\documentclass[10pt]{iopart}

\usepackage{epsfig}
\usepackage{graphics}
\usepackage{graphicx}
\usepackage{bm}
\usepackage{color}
\usepackage{hyperref}
\usepackage{amssymb}
\usepackage{longtable}
\usepackage{rotating}
\usepackage{color}
\usepackage{epsf}

\def\bea{\begin{eqnarray}}
\def\eea{\end{eqnarray}}
\newcommand{\be}{\begin{equation}}
\newcommand{\ee}{\end{equation}}
\newcommand{\ber}{\begin{eqnarray}}
\newcommand{\eer}{\end{eqnarray}}

\newcommand{\ie}{\emph{i.e.} }
\newcommand{\eg}{\emph{e.g.} }

\def\cross{\times}

\def\bea{\begin{eqnarray}}
\def\eea{\end{eqnarray}}
\newcommand{\beq}{\begin{equation}}
\newcommand{\eeq}{\end{equation}}
\newcommand{\beast}{\begin{equation*}}
\newcommand{\eeast}{\end{equation*}}

\newcommand{\lambdav}{\vec \lambda}

\newcommand{\betav}{\vec \beta}

\newcommand{\nDR}{n_{\textsc{dr}}}
\newcommand{\dx}{{\rm d}x}
\newcommand{\mean}{\mathrm{mean}}

\def\mHz{\mathrm{mHz}}
\def\nHz{\mathrm{nHz}}

\def\A{\mathcal{A}}

\begin{document}

% Title

\title[Delayed Rejection Markov chain Monte Carlo approach for LISA analysis]
{Studying stellar binary systems with the Laser Interferometer Space Antenna using Delayed Rejection Markov chain Monte Carlo methods}

\author{Miquel~Trias$^1$, Alberto~Vecchio$^2$ and John~Veitch$^2$}
\address{$^1$ Departament de F\'{\i}sica, Universitat de les Illes
Balears, Cra. Valldemossa Km. 7.5, E-07122 Palma de Mallorca, Spain}
\address{$^2$ School of Physics and Astronomy, University of Birmingham, Edgbaston, Birmingham B15 2TT, UK}
\eads{\mailto{miquel.trias@uib.es}, \mailto{av@star.sr.bham.ac.uk}, \mailto{jveitch@star.sr.bham.ac.uk}}

%%%%%%%%%%%%%%%%%%%%%%%%%%%%%%%%%%%%%%%%%%%%%%%%%%%%%%%%%%%%%%%
\begin{abstract}
Bayesian analysis of LISA data sets based on Markov chain Monte Carlo methods has been shown to be a challenging problem, in part due to the complicated structure of the likelihood function consisting of several isolated local maxima that dramatically reduces the efficiency of the sampling techniques. Here we introduce a new fully Markovian algorithm, a Delayed Rejection  Metropolis-Hastings Markov chain Monte Carlo method, to efficiently explore these kind of structures and we demonstrate its performance on selected LISA data sets containing a known number of stellar-mass binary signals embedded in Gaussian stationary noise. 
\end{abstract}
\date{\today}


%%%%%%%%%%%%%%%%%%%%%%%%%%%%%%%%%%%%%%%%%%%%%%%%%%%%%%%%%%%%%%
\section{Introduction}
\label{s:intro}

Surveys of our galaxy in the gravitational-wave band with the Laser Interferometer Space Antenna, LISA \cite{LISA-Pre-Phase} will yield the largest sample of close stellar-mass binary systems, in particular white dwarfs. LISA is in fact expected to individually resolve $\sim 10^4$ stellar-mass binary systems that are sufficiently bright to stand above the confusion noise produced by the whole population of binaries, most of which are unresolved~\cite{Nelemans:2001,Farmer:2003pa}. Gravitational waves from stellar-mass compact objects in the low-frequency observational window are long lived, and in order to extract the maximum amount of information, one needs to be able to disentangle and accurately measure the parameters of a large number of sources overlapping in time and frequency space. This represents a challenging data analysis problem, whose successful solution is essential for the full science exploitation of the LISA data set.

Due to the large number of sources to be analysed at the same time and the fact that this number is unknown, Markov chain Monte Carlo (MCMC) techniques \cite{Cornish:2005qw, Crowder:2006wh, Crowder:2006eu, Crowder:2007ft, Stroeer:2007tg, TVV:2008} and their extension to Reversible Jump Markov chain Monte Carlo (RJMCMCs) \cite{Umstatter:2005jd, Cornish:2007if,Littenberg:2009} are expected to be one of the most powerful search methods, although analysis approaches based on matched-filtering with a template bank have also been explored~\cite{PrixWhelan:2007,WhelanPrixKhurana:2008}.

In MCMC applications to LISA data analysis, one of the key problems that one needs to tackle is the fact that the joint posterior probability density function (PDF) of the source parameters -- the target distribution -- presents a multimodal structure, often with secondary maxima well separated from the mode, see Figure~\ref{f:likelihood} for an example. In such situations, maintaining a high acceptance ratio while sampling the full structure of the posterior PDF becomes very difficult, and may lead to a negligible efficiency of the algorithm. Several solutions have been proposed to increase the mixing and exploration ability of the chains, such as using parallel chains exploring the parameter space that, at a certain point, can be swapped (\eg Metropolis-coupled MCMC algorithms \cite{Geyer:1991}), or other methods like simulated annealing \cite{Kirkpatrick:1983} or simulated tempering \cite{Marinari:1992, Geyer:1995} that consist in adding a `temperature' factor to the target distribution in order to make it smoother at the beginning and therefore to promote the movement of the chain. Examples of the use of some of these techniques to compute the posterior PDF for white dwarf binaries in LISA data sets are given in \emph{e.g.} Refs.~\cite{Cornish:2005qw,Crowder:2006wh,Crowder:2006eu,Cornish:2007if,Littenberg:2009}. However, as the number of dimensions increases and the target distribution becomes more complicated, it is not clear which technique works best. For instance, running $N$ parallel chains requires $N$ times more computing time; moreover, it has been shown that in some cases, this approach may yield low efficiency~\cite{Cornish:2008}. Non-Markovian explorations during an initial ``search phase'' have been considered to quickly identify the neighbourhood of the main mode of the distribution; however these strategies suffer from the significant draw-back that they cannot be readily extended to the relevant case of an unknown number of dimensions (\emph{i.e.} unknown number of signals) and as a consequence model selection. For all these reasons, it is useful to explore alternative approaches.

In this paper we explore a different strategy for a fully Markovian sampling of multimodal posterior PDFs: a \emph{Delayed Rejection} (DR) Metropolis-Hastings Markov chain Monte Carlo method. Delayed Rejection was originally introduced by Tierney and Mira~\cite{TierneyMira:1999,Mira:2001} for fixed dimension problems and then extended by Green and Mira~\cite{GreenMira:2001} to transdimensional problems. The algorithm is now found in a wide variety of science applications, see \emph{e.g.} Refs.~\cite{Harkness:2000, Robert:2003, Al-Awadhi:2004, Umstatter:2004, Raggi:2005, Haario:2006}. In general, in all applications so far, either the number of steps was very small (2 or 3) or the assumption of symmetrical proposals is made. We have recently extended the DR algorithm to its most general scheme, using an arbitrary number of stages~\cite{TVV:2009}. Here we demonstrate its power in applications to the LISA white dwarf problem, by applying it to the analysis of data sets in the restricted case in which the dimensionality of the problem (that is proportional to the number of white dwarf presents in the data set) is fixed and known.

We would like to emphasise that although the waveforms that we consider here are the simplest present in the LISA data set, this specific example tackles all the conceptual issues that will be present in other LISA data analysis problems involving more complex signals, such as those from spinning massive black hole binaries and extreme-mass ratio inspirals. The DR scheme will therefore be applicable to the development of MCMC analysis algorithms targeted at the study of those sources.

% --------------------

\section{Delayed Rejection Markov chain Monte Carlo methods}
\label{s:drmcmc}

\subsection{The general method}
\label{ss:method}

The Metropolis-Hastings algorithm is widely used in applications of Markov chain Monte Carlo methods to sample a \emph{target distribution} $\pi(x)$, in our case the posterior PDF of a set of parameters.  Given a state of the Markov chain $\lambdav$, one proposes a new state $\lambdav'$  drawn from a \emph{proposal distribution} (or transition kernel) $q(\lambdav, \lambdav')$, the probability of $\lambdav'$ given $\lambdav$. This new state is accepted with probability
\beq\label{e:alpha}
	\alpha(\lambdav , \lambdav') = 1 \wedge \left\lbrace \frac{\pi(\lambdav') \, q(\lambdav' , \lambdav)}
							{\pi(\lambdav) \, q(\lambdav , \lambdav') }\right\rbrace \; ,
\eeq
and the chain remains at $\lambdav$ with probability $1 - \alpha(\lambdav , \lambdav')$. In the previous equation, the notation $a \wedge b$ (for any real numbers $a$ and $b$) stands for the minimum between $a$ and $b$, and as a consequence the acceptance probability is limited to unity in the case where the ratio is $>1$.

An effective way to improve the efficiency of the MCMC algorithm has been proposed in the form of Delayed Rejection~\cite{TierneyMira:1999, Mira:2001, GreenMira:2001}, and now we briefly summarise it. Suppose that the current state of the chain is $\lambdav$. A candidate, $\betav_1$, is generated from a proposal distribution $q_1(\lambdav , \betav_1)$ and accepted with probability
\beq
	\alpha_1(\lambdav , \betav_1) = 1 \wedge \left\lbrace \frac{\pi(\betav_1) \, q_1(\betav_1 , \lambdav)}
	 																					     {\pi(\lambdav) \, q_1(\lambdav , \betav_1) }\right\rbrace \; ,
\eeq
as in a standard Metropolis-Hastings algorithm, see Equation~(\ref{e:alpha}). If $\betav_1$ is not accepted, instead of rejecting it, one can use this information and propose a new state $\betav_2$ from a (in principle different) proposal distribution $q_2(\lambdav , \betav_1 , \betav_2)$ which may use information about the rejected state $\betav_1$. 

Moreover, in general the $i-$th stage of the DR chain is as follows~\cite{TVV:2009}: if the state $\betav_{i-1}$ is proposed and rejected, one can propose a new candidate $\betav_i$ from the proposal $q_i(\lambdav , \betav_1 , \ldots , \betav_i )$ and accept it with probability
\bea \label{eq:mastereq}
\alpha_i (\lambdav , \betav_1, \ldots , \betav_i) & = &
1 \wedge \left\lbrace \frac{\pi(\betav_i)~q_1(\betav_i,\betav_{i-1})~q_2(\betav_i,\betav_{i-1},\betav_{i-2})~\ldots ~
q_i(\betav_i,\betav_{i-1}, \ldots , \lambdav )}
{\pi(\lambdav )~q_1(\lambdav ,\betav_1)~q_2(\lambdav ,\betav_1,\betav_2)~\ldots ~q_i(\lambdav ,\betav_1, \ldots , \betav_i)} \right.
\nonumber\\
& & \hspace{-1.5cm} \left. \frac{\left[ 1-\alpha_1(\betav_i,\betav_{i-1}) \right] \left[ 1-\alpha_2(\betav_i,\betav_{i-1},\betav_{i-2}) \right] 
\ldots \left[ 1-\alpha_{i-1}(\betav_i, \ldots \betav_1) \right] }
{\left[ 1-\alpha_1(\lambdav ,\betav_1) \right] \left[ 1-\alpha_2(\lambdav ,\betav_1,\betav_2) \right] \cdots 
\left[ 1-\alpha_{i-1}(\lambdav , \ldots \betav_{i-1}) \right] } \right\rbrace \; .
\eea
Notice that with the notation we are using, $q_i(\lambdav ,\betav_1, \ldots , \betav_{i-1} , \betav_i)$ represents the proposal probability of $\betav_i$ given $\{ \lambdav , \betav_1 , \ldots , \betav_{i-1} \}$ where the order of the parameters does matter. By imposing detailed balance at each stage and deriving the acceptance probability that preserves it, the resulting chain will be a reversible Markov chain with invariant distribution $\pi$.

From a theoretical point of view, it can be demonstrated in general (see Sec.~2.2 of Ref.~\cite{TVV:2009}) that the variance of an estimate made from a chain using Delayed Rejection is always smaller than that produced with a standard Markov chain. Moreover, besides increasing the efficiency of the sampling, this method allows us to propose a new candidate using the information of the past elements of the chain. We will profit from these two properties in increasing the efficiency of the computation of the marginalised posterior distributions generated by integrating multimodal likelihood functions.

% ----

\subsection{Key rules on the choice of proposals}
\label{ss:key_rules}

An essential point of the DR scheme is that Eq.~(\ref{eq:mastereq}) has been derived by imposing that the backward path, from $\betav_i$ to $\lambdav$, follows the forward path, from $\lambdav$ to $\betav_i$, in reverse order. As a consequence, this property forces one to choose proposal distributions that preserve, in some way, the reversibility of the chain in order to avoid negligible acceptance probabilities. Full details about the general implementation of the algorithm are provided in Ref.~\cite{TVV:2009}; here we focus on the choice of proposals relevant to searches for white dwarf binary systems.

When we run a Metropolis-Hastings MCMC chain -- likely exploring a secondary mode of the distribution -- we can start a DR chain with a large jump that attempts to reach the mode of the distribution; such transition, likely to be rejected but hopefully closer to the mode, is followed by many small ones in order to explore the new region of the parameter space. 
In terms of the proposal PDFs that appear in Eq.~(\ref{eq:mastereq}), this translates into having $q_1(\lambdav, \betav_1) \equiv q_a(\lambdav , \betav_1)$ mainly proposing large transitions in the parameter space, and $q_j(\lambdav , \betav_1 , \ldots , \betav_j) \equiv q_b(\bar{x}[\lambdav , \ldots , \betav_{j-1}] , \betav_j)$ for $j \geq 2$, proposing small transitions. This choice inevitably adds some degree of non-reversibility that significantly degrade the  acceptance probabilities, see Eq.~(\ref{eq:mastereq}). One could however achieve an efficient DR algorithm, by using proposals that allow for both small and big jumps, such as a mixture of three Gaussians (3-Gaussian, from now on) symmetrical functions:
\begin{equation}
q_{a, b} (\bar{x},x) = N_{a, b} ~G(\bar{x} , \sigma_1 ; x) + \frac{1-N_{a, b}}{2} \left[ G(\bar{x}-\mu , \sigma_2 ; x) + G(\bar{x}+\mu , \sigma_2 ; x) \right] \,;
\end{equation}
in the previous equation, $\bar{x}$ may be a function of the old elements of the DR chain,  $N_{a, b}$ represents the probability of proposing a value from the central Gaussian and $G(\mu , \sigma ; x) \equiv \frac{1}{\sqrt{2 \pi} \sigma} \exp \left[ -\frac{(x-\mu)^2}{2 \sigma^2} \right]$. Notice that an efficient exploration of multimodal structures could in principle be achieved by taking $N_a \to 0$ and $N_b \to 1$, but this in general would lead to negligible acceptance probabilities. In addition, the central value of the 3-Gaussian, $\bar{x}[\lambdav , \ldots , \betav_{j-1}]$ that is computed at each DR stage making use of the information from the old elements of the chain, should be independent of the particular sample of the chain. In our application we use the mean of all the past elements of the chain after the big jump,
\beq \label{eq:def_central_next_proposal}
\bar{x} [\lambda , \beta_1 , \ldots , \beta_{j-1}] \equiv \mean (\beta_1 , \ldots , \beta_{j-1})\,.
\eeq

With these choices, all the proposals used during the DR are characterised by $6$ parameters: $\{ \sigma_1 , \sigma_2 , \mu , N_a , N_b\}$ and the maximum number of elements in a DR chain, $\nDR$; as a consequence, the contribution of the ratio of proposals to the final acceptance probability, Eq.~(\ref{eq:mastereq}), can be quantified (see Figs.~4 and 6 of \cite{TVV:2009}) in terms of these parameters. The choice of these parameters is the essential step in obtaining a DR algorithm that probes in an efficient way the posterior PDF. In practice, they can be chosen as follows: (i) $\{ \sigma_1 , \sigma_2 , \mu \}$, are chosen to have a proposal adapted to the multimodal structure of the target distribution, $\mu$ being the typical distance between two maxima of the target distribution and $\sigma_i$ their typical width; as consequence a qualitative understanding of the typical scales in parameter space of the likelihood function is important (we deal in detail with this point in the next Section); (ii) from $\{ \sigma_1 , \sigma_2 , \mu \}$, one can then identify the panels from Figs.~4 and 6 of \cite{TVV:2009} corresponding to the closest $\sigma_1 / \mu - \sigma_2 / \mu$ values; first, one looks at the \emph{losses due to the central proposal evolution (CPE)} (Figure~6 of \cite{TVV:2009}) to select $N_b$ high enough\footnote{The tolerance range depends on the typical ratio between the likelihood of the main maximum and the likelihood of the secondary ones. In our experience, allowing losses of the order of $e^{-3} - e^{-4}$ is acceptable.} to have acceptable expected values of this ratio of proposals; as one normally works with $\nDR \gtrsim 100$, the results depend very weakly on $\nDR$; (iii) one then needs to consider the \emph{losses due to an asymmetrical proposal (AP)} (Figure~4 of \cite{TVV:2009}) in order to choose $N_a$; (iv) finally, as again the dependency on $\nDR$ is weak, its choice will be mainly driven by the constraints on the computational costs.

So, by following the prescriptions described here and having some previous knowledge about the multimodal structure of the target distributions that has to be sampled, one can implement a DR MCMC algorithm with an arbitrary numer of steps, $\nDR$, capable of efficiently sampling these multimodal distributions. The algorithm is completely general, and can be applied to a wide range of problems. In the following sections we will focus on the application of the DR MCMC algorithm to the search of a known number of stellar-mass binary signals in Gaussian and stationary noise with LISA.

% --------------------

\section{Application to LISA observations of white dwarfs}
\label{s:application}

We consider a LISA data set -- that we indicate with $d_{A,E}$, to highlight the use of the two noise-orthogonal Time-Delay-Interferometry observables $A$ and $E$, see \emph{e.g.} Ref.~\cite{tdi-lrr} for a review -- containing a known number, $N$, of signals from stellar-mass binary systems ($h_{A,E}$) buried in Gaussian stationary noise ($n_{A,E}$):
\beq
d_{A,E}(t) = n_{A,E}(t; \vec{\zeta}_\mathrm{n}) + h_{A,E}(t;\vec{\zeta}_\mathrm{s})\,.
\label{e:da}
\eeq
The vector $\vec{\zeta} = \{\vec{\zeta}_\mathrm{n}, \vec{\zeta}_\mathrm{s}\}$ contains all the $M$ unknown parameters that describe the problem, where $\vec{\zeta}_\mathrm{n}$ and $\vec{\zeta}_\mathrm{s}$ are the noise and signal parameters, respectively. For the specific case considered here, we focus on monochromatic (in the source reference frame) white dwarfs and purely Gaussian and stationary (with zero mean and variance $\sigma_n^2$) instrumental noise. The total number of parameters that describe the problem -- \ie the dimensionality of $\vec{\zeta}$ -- is $M = 1 + 7 N$
\bea
\vec{\zeta}_\mathrm{s} & = & \left\lbrace \{f_0, \A_+, \frac{\A_\cross}{\A_+} , \phi, \theta, \psi, \varphi_0\}_1, \{f_0, \A_+ , \frac{\A_\cross}{\A_+} , \phi, \theta, \psi,  \varphi_0\}_2 , \ldots \right\rbrace \,,
\nonumber\\
\vec{\zeta}_\mathrm{n} & = & \sigma_n^2\,,
\eea
where $f_0$ is the constant frequency of the signal (with respect to the Solar System Barycentre), $\A_+$ and $\A_\cross$ are the amplitudes of the two GW polarizations, the longitude $\phi$ and co-latitude $\theta$ are the two angles that define the source's sky location, $\psi$ is the polarization angle of the gravitational wave and $\varphi_0$ is a constant that fixes the initial phase of the signal.

We make use of the Delayed Rejection MCMC algorithm introduced in the previous Section (see~\cite{TVV:2009} for more details) to compute the (marginalised) posterior PDFs of the noise and signal parameters. By applying the Bayes' theorem, we can compute the \emph{joint} posterior density function, $p(\vec{\zeta}| d)$, of $\vec{\zeta}$ given the data sets, $d$,
\beq
p(\vec{\zeta}|d) = \frac{p(\vec{\zeta}) \, \mathcal{L}(d | \vec{\zeta})}{p(d)}\, ,
\label{e:pdf}
\eeq
where $\mathcal{L}(d | \vec{\zeta})$ is the likelihood function, $p(\vec{\zeta})$ the prior probability density function, and $p(d)$ is the marginal likelihood. From a Markov chain Monte Carlo algorithm perspective, the target distribution that needs to be sampled is $\pi(\vec{\zeta}) \equiv p(\vec{\zeta}) \, \mathcal{L}(d | \vec{\zeta}) \propto p(\vec{\zeta}|d)$. Since the resulting chain is a Markov chain with invariant distribution $\pi$, the average along a sample path of a function $f$ is an asymptotically unbiased estimate of $\int f(x) \pi(x) \dx$; in particular, the marginalised posterior PDF over a set of parameters, can be directly extracted from the output Markov chain of the algorithm.

% ----

\subsection{Structure of a single source target distribution}
\label{ss:structure_target}

\begin{figure}
\begin{center}
\begin{tabular}{cc}
\multicolumn{2}{c}{$f \simeq 1 \mHz$} \\
\includegraphics[width=0.48\textwidth]{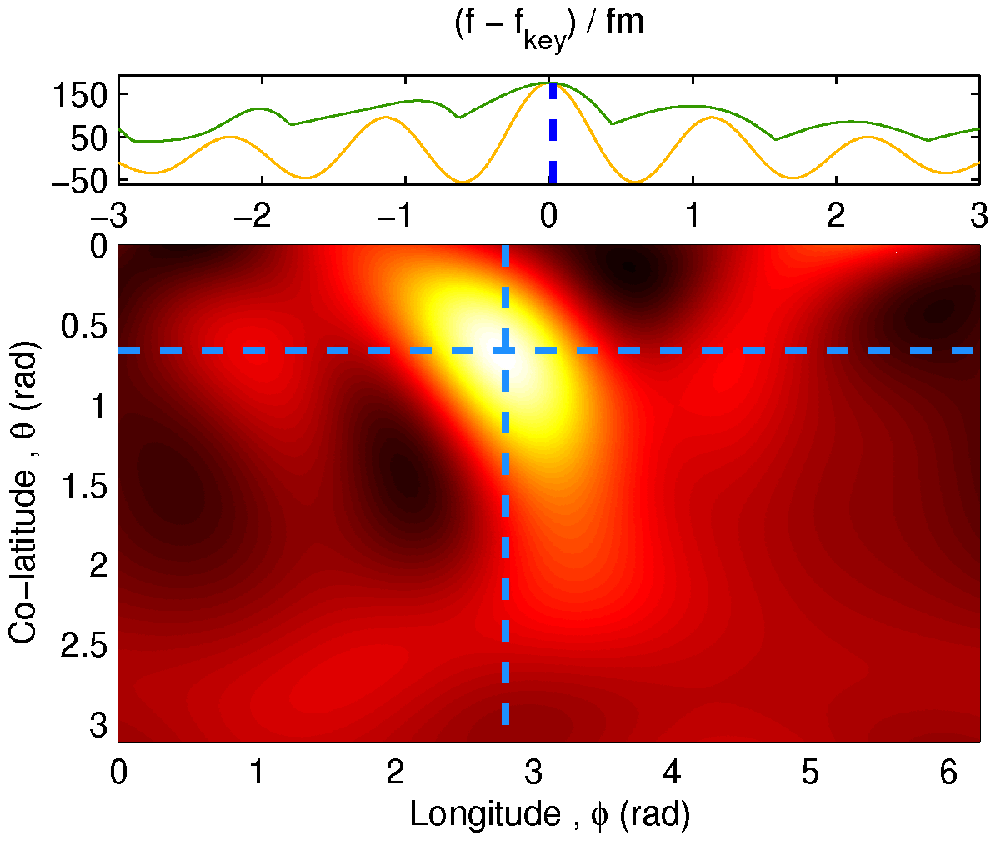} & \includegraphics[width=0.48\textwidth]{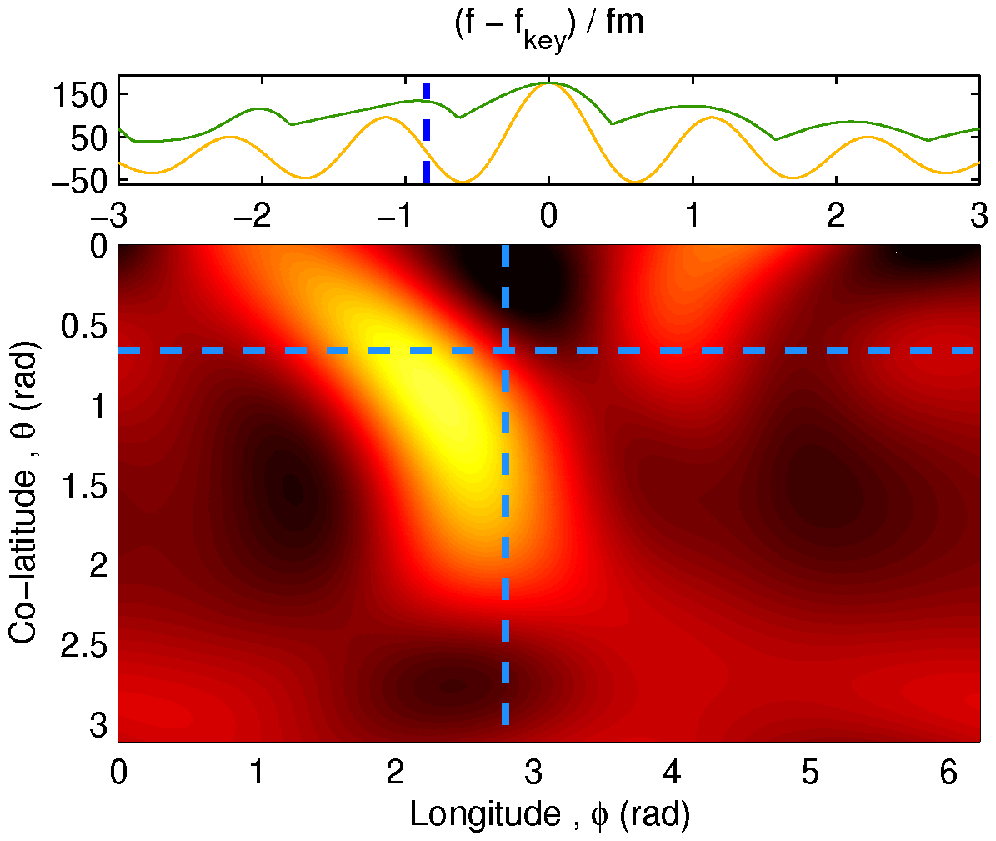} \\ \\
\multicolumn{2}{c}{$f \simeq 3 \mHz$} \\
\includegraphics[width=0.48\textwidth]{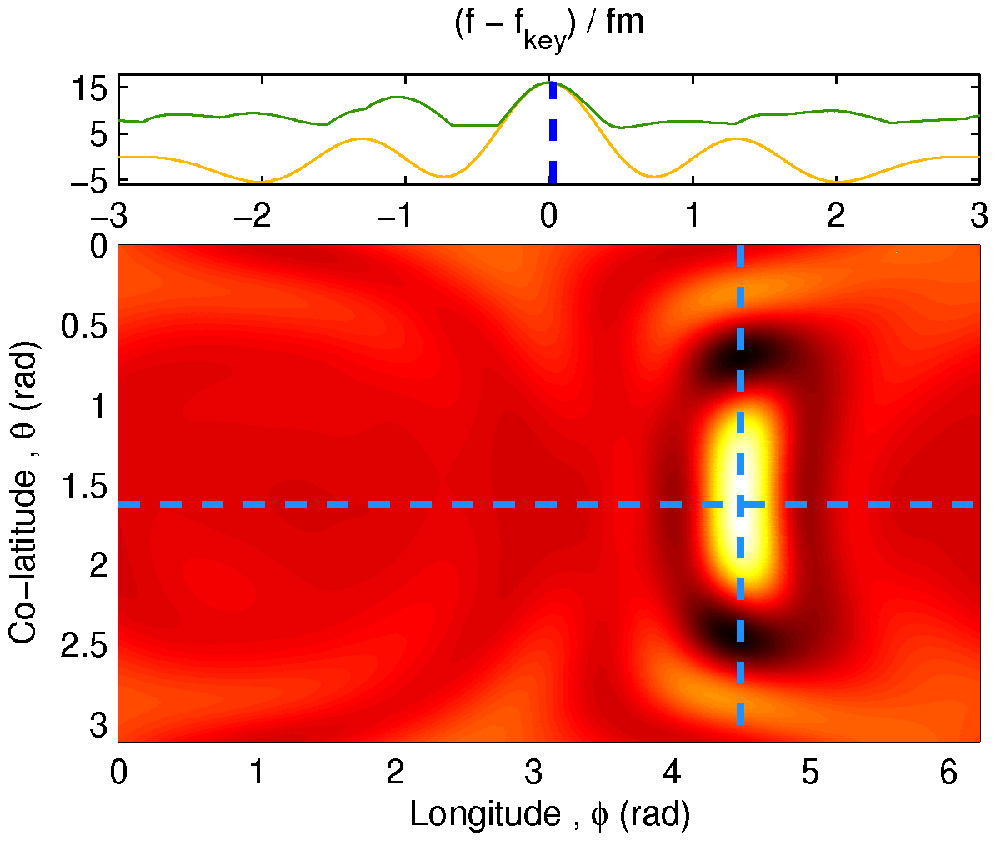} & \includegraphics[width=0.48\textwidth]{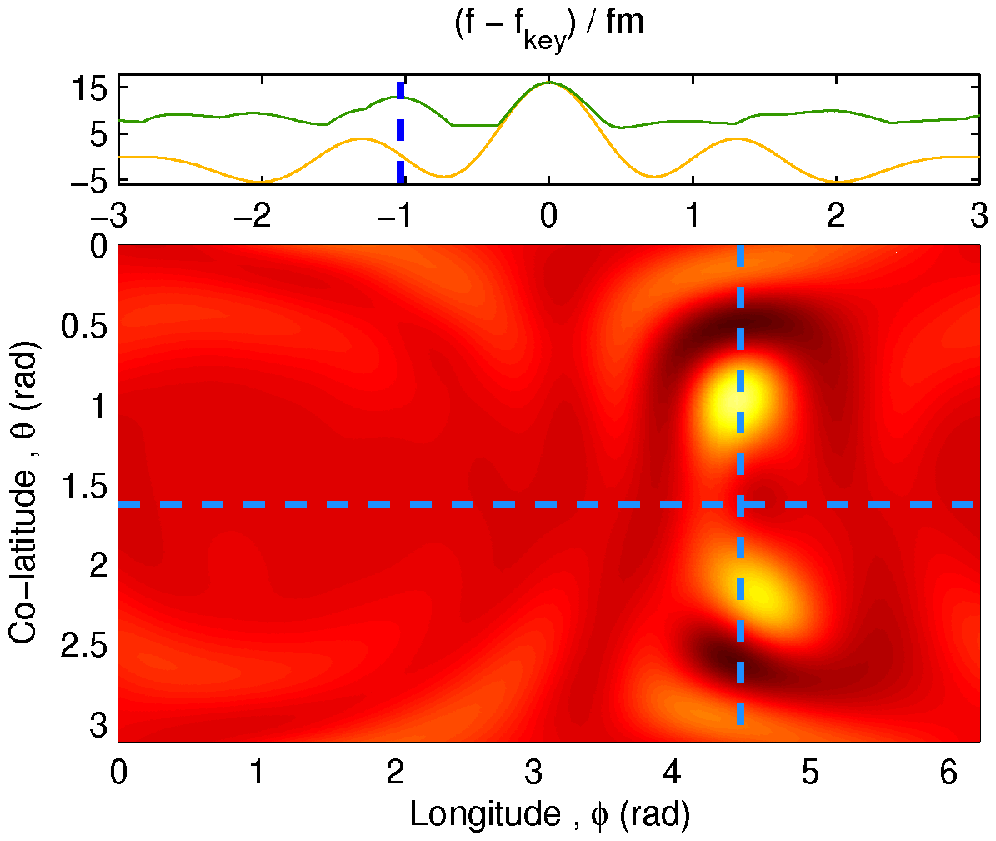} \\ \\
\multicolumn{2}{c}{$f \simeq 10 \mHz$} \\
\includegraphics[width=0.48\textwidth]{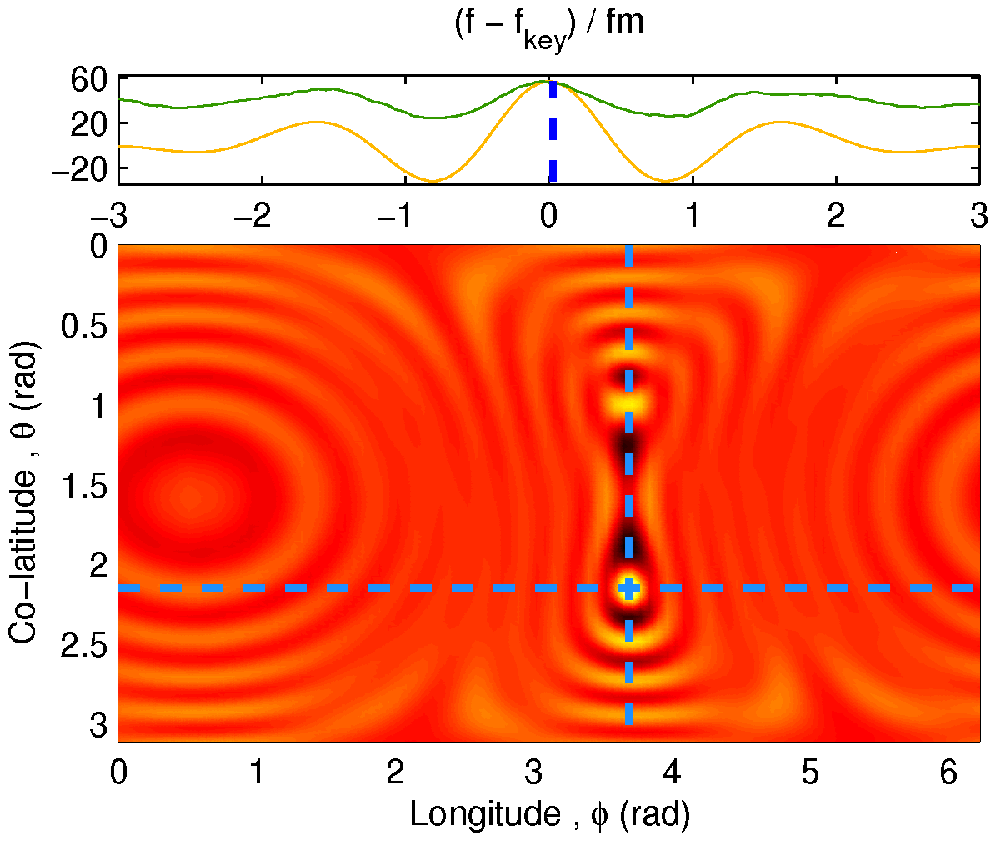} & \includegraphics[width=0.48\textwidth]{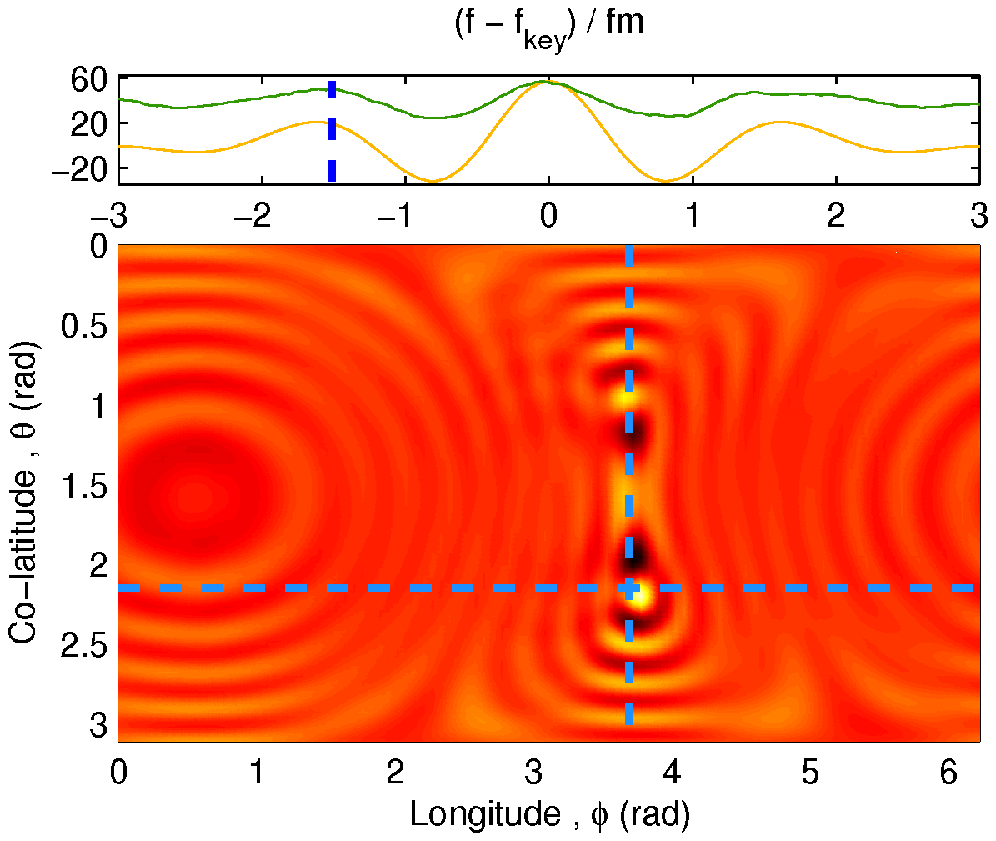}
\end{tabular}
\end{center}
\caption{Colour maps with the marginalised likelihood function in the co-latitude ($\theta$) -- longitude ($\phi$) plane at two particular frequency values corresponding to the main (left-hand plots) and secondary maxima (right-hand plots); and, linear plots in the top panels with the marginalised (lower solid line) and maximised (higher solid line) likelihood as a function of the difference between the signal's frequency, $f$, and its actual value reported in the key file, $f_{key}$, measured in units of the LISA's modulation frequency, $f_m = \,1\mathrm{yr}^{-1}$. These plots were repeated for three frequencies. The dashed lines in the colour maps represent the actual sky location of the source, meanwhile the dashed line in the top panels represents the frequency value where the colour plots were generated.}
\label{f:likelihood}
\end{figure}

\begin{figure}
\begin{center}
\begin{tabular}{ccc}
\includegraphics[width=0.35\textwidth]{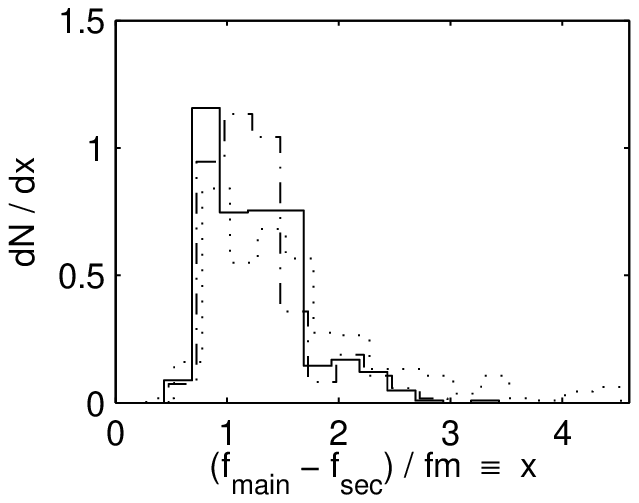} & \hspace{-0.75cm}
\includegraphics[width=0.35\textwidth]{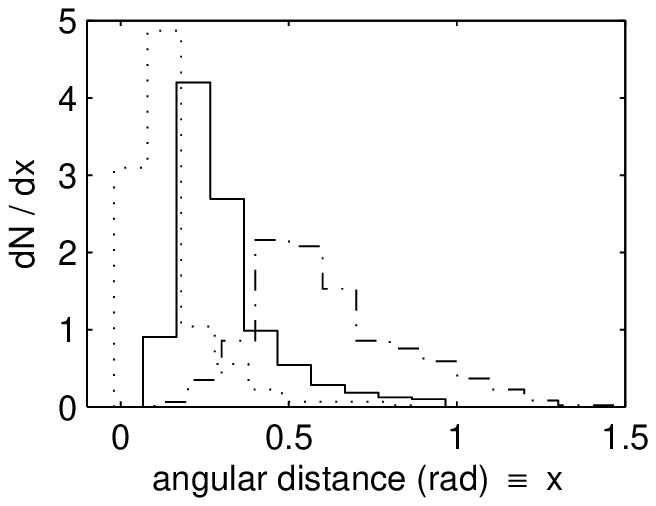} & \hspace{-0.75cm}
\includegraphics[width=0.35\textwidth]{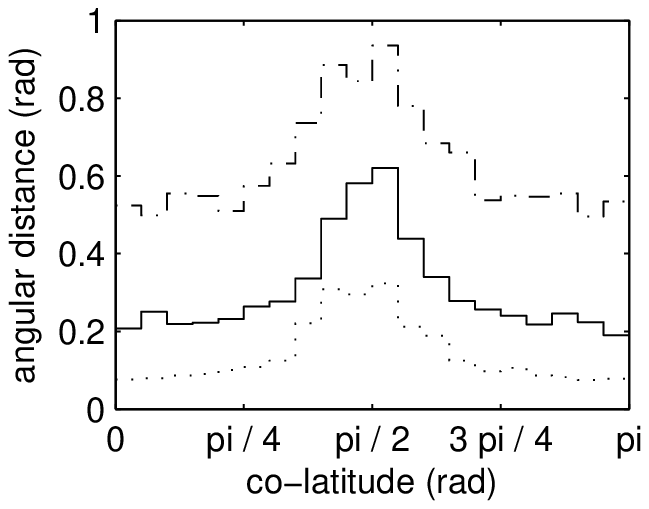} 
\end{tabular}
\end{center}
\caption{Results obtained after performing $500$ MCs over all the $7$ parameters that characterise the GW signal measured from a stellar-mass binary system and studying the relative position of the secondary maxima with respect to the main one. The frequency values are restricted to narrow bands around $1 \mHz$ (dash-dot), $3 \mHz$ (solid) and $10 \mHz$ (dotted). (a) The left-hand plot shows the distributions of relative distance in frequency between the two maxima and in units of the LISA's modulation frequency, $f_m = 1 yr^{-1}$ ; (b) the central panel plots the distributions of the spherical angular distance ; (c) the right-hand plot shows the dependency of these averaged values of angular distance with the co-latitude of the source.}
\label{f:SecMax}
\end{figure}

%%%

The application of the DR algorithm requires some previous knowledge of the multimodal structure present in the target distribution of the problem at hand. This structure is intrinsic to the single source posterior PDF and, in particular, to its likelihood function, $\mathcal{L}(d | \vec{\zeta})$. Local maxima \cite{Crowder:2006eu} occur in the likelihood surface at approximately multiples of the modulation frequency ($f_m = 1\,\mathrm{yr}^{-1}$) around the frequency of the signal measured at the Solar System Barycentre. Although the main multimodal structure occurs in the frequency parameter, other parameters, like the sky position, are correlated to the frequency, and this is reflected in the structure of secondary maxima of the likelihood function.

In Fig.~\ref{f:likelihood} we explore the structure of the likelihood function in the 3-dimensional space $\{f , \theta , \phi \}$, around the global maximum and the secondary mode, for three different frequency values. It can be observed how local modes of the maximised likelihood function appear around the actual frequency value at distances that are multiples of $f_m$ and, at the same time, the sky location is affected. As the signal frequency increases, a smaller change in the sky location can produce the same frequency shift due to the Doppler effect produced by the LISA motion, which explains the finer structures in the panels at the bottom of Fig.~\ref{f:likelihood}. In order to have a more systematic and quantitative understanding of the multimodal structure of the likelihood function, we have performed $500$ Monte Carlo (MC) simulations around different frequencies, taking random values for all the $7$ parameters that characterise a single white dwarf binary signal, and have computed the relative position (in parameter space) of the secondary maxima with respect to the main mode. The results of this Monte Carlo's are summarised in Fig.~\ref{f:SecMax}; from those we can make a suitable choice of the parameters $\{ \sigma_1 , \sigma_2 , \mu \}$ that determine the proposals of the DR stage.

Some general conclusions, that are important for the tuning of the DR algorithm can be drawn. The frequency separation between the two modes appears to be independent of the actual frequency of the signal and the values are mostly concentrated between $0.75 f_m$ and $1.75 f_m$. Moreover, we can estimate the angular distance on the two-sphere, $\Delta \hat{\Omega}$, between the sky position corresponding to the main and the secondary maximum. From the central plot of Fig.~\ref{f:SecMax} we observe that $\Delta \hat{\Omega}$  progressively shifts to lower values as the signal's frequency increases, and that (right-hand plot of Fig.~\ref{f:SecMax}) the mean value of $\Delta \hat{\Omega}$ changes with the co-latitude, $\theta$, producing secondary maxima that are more separated near the ecliptic plane than at the poles. A qualitative understanding of the dependency observed in Figure~\ref{f:SecMax} can be simply derived from the expression of the Doppler shift, $\dot{\phi}_{D , i} (t) = {2 \pi f_0}/{c} ~\hat{k} \cdot \vec{v}_i (t)$, where $\hat{k}$ is the unit vector from the source to the detector and $\vec{v}_i$ is the velocity vector of the $i$-th LISA spacecraft. From here, we observe that the required changes in the sky location needed to obtain a certain frequency shift, will be inversely proportional to $f_0$.

% ----

\subsection{Parameters for the Delayed Rejection proposal distributions}
\label{ss:proposal_params}

The multimodal structure of the target distribution relevant for a search for stellar-mass binaries in the LISA data is mainly present in the frequency of the signal, although there are correlations with the sky location angles. The solution that we have adopted here, as a first implementation of the DR algorithm, consists in applying a scheme where only these three parameters, $\{ f , \theta , \phi \}$, are updated: the first one, $f$, using a mixture of three Gaussian distributions and the two angles defining the sky location drawn from a single Gaussian proposal centered at the previous element of the DR chain and width according with the results obtained in Fig.~\ref{f:SecMax}. With this, only the frequency proposals can contribute with small factors to the acceptance probability of Eq.~(\ref{eq:mastereq}), so we will now focus on these proposals.

By following the prescription given in Sec.~\ref{ss:key_rules}, we select the first three parameters, $\{ \sigma_1 , \sigma_2 , \mu \}$, that characterise the 3-Gaussian distribution for the frequency. In light of the results shown in Figs.~\ref{f:likelihood} and \ref{f:SecMax}, our proposals can have the modes separated by $\mu = 1.25 f_m$, choosing a $\sigma_1 = f_m/2$ wide central Gaussian when exploring local modes and $\sigma_2 = 0.2 f_m$ for the external Gaussians. This means that we propose from the external Gaussians, $95.4\%$ of the values in the $(0.85 f_m , 1.65 f_m)$ interval.

Moreover, by looking at the corresponding panels of Figures~6 and 4 of \cite{TVV:2009} (in our case, $\sigma_1 / \mu = 0.4$ and $\sigma_2 / \mu = 0.16$), one can read out the values for $N_b$ and $N_a$. Finally, the maximum number of iterations allowed in a Delayed Rejection is fixed mainly for computational reasons; from our experience we think that $\nDR = 1000$ to $3000$ are adequate values, depending on whether one prefers to attempt more frequent short DR chains, or longer chains less frequently. In summary, a DR algorithm for efficiently explore the joint posterior PDF in the frequency space can be defined by the following parameters:
\begin{equation} \label{e:paramsDR}
\{ \sigma_1 , \sigma_2 , \mu , N_a , N_b , \nDR \} = \{0.5 f_m , 0.2 f_m , 1.25 f_m , 0.15 , 0.95 , 1000 \}
\end{equation}

% ----

\subsection{Results}
\label{ss:results}

\begin{figure}
\includegraphics[width=\textwidth]{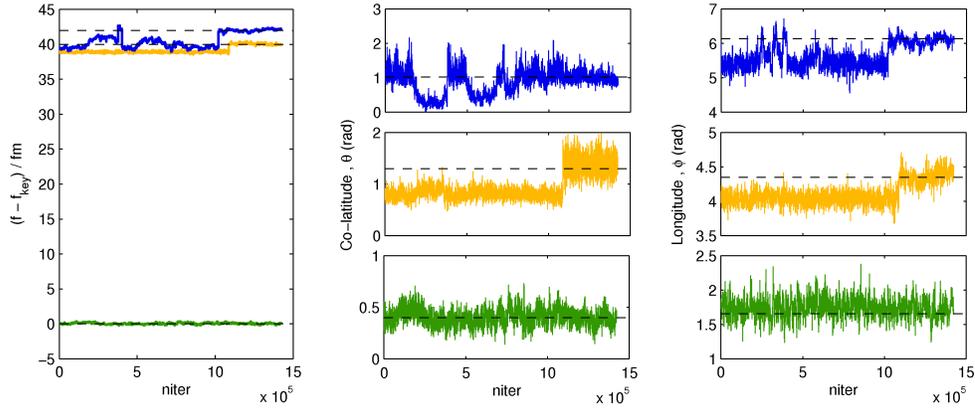}
\caption{Output chains obtained after running the DR MCMC algorithm presented in this article over a LISA data set containing $3$ sources around $1.1 \mHz$ buried in Gaussian stationary noise, two of them separated in frequency by $2 f_m$, $f_m$ being the frequency modulation of LISA, and the third one located at $40 f_m$ from those two. All the other parameters are chosen randomly. The dashed lines represent the actual values of the parameters for each source and we use solid lines of different colours to represent the parameters from different chains.}
\label{f:Example_3srcs}
\end{figure}

\begin{figure}
\includegraphics[width=\textwidth]{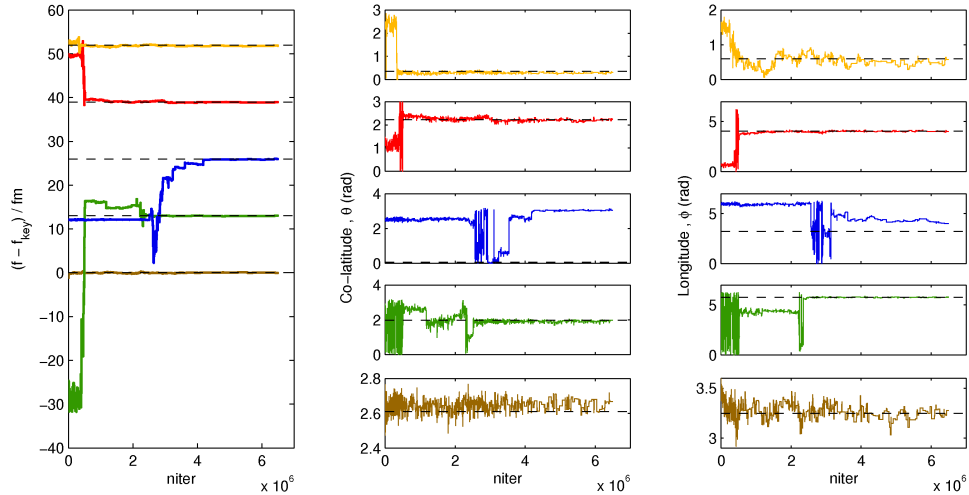}
\caption{The same plots as in Fig.~\ref{f:Example_3srcs} but here working with a $5$ sources data set around $1.6 \mHz$, with a separation of $13 f_m$ between each source. Only the third chain (blue in the on-line version) converged to a wrong sky location (the opposite to the actual one) because of degeneracies present in the likelihood surface and its proximity to one of the poles.}
\label{f:Example_5srcs}
\end{figure}

We have implemented a DR MCMC algorithm to search for a known number of stellar-mass binary signals in LISA data, assuming Gaussian and stationary noise. In this section we provide example results obtained by applying the analysis method on two different 1-yr long data sets: (i) one containing three sources at $1.1~\mHz$ separated in frequency by $2 f_m$ and $40 f_m$, see Fig.~\ref{f:Example_3srcs}, and (ii) the other including $5$ sources around $1.6~\mHz$ equally separated in frequency by $13 f_m$, see Fig.~\ref{f:Example_5srcs}. From these figures we can see that after the burn-in period -- we note that the DR part is switched on only after burn-in -- that has been discarded, most of the chains are sampling either the actual mode of the signals or secondary maxima. At this point, the Delayed Rejection is switched on (with probability $1/3000$) and from time to time a new region $\approx 1 f_m$ away from the current position of the chain, is explored searching for higher maxima. We observe that once the chain finds the right mode in the frequency space, all the other parameters also move their actual values.

With the DR, we are drawing $\nDR = 1000$ proposals mainly targeted to a different mode of the posterior PDF with a probability of $1/3000$, most of which are rejected. By keeping the same odds with a standard MCMC, we would have the same ability of exploring different maxima but this algorithm would yield an extremely correlated Markov chain, \ie greater variances of the estimates once it has reached the stationary distribution. On the other hand, one could use a standard MCMC proposing big jumps in the parameter space with probability $1/3000$, and the resulting Markov chain would have the same autocorrelation as the one obtained from a DR MCMC; however, the ability to explore different maxima of the target distribution would decrease by a factor $\nDR = 1000$. We refer the reader to Ref. \cite{TVV:2009} for further discussions and examples.

The results of Figs.~\ref{f:Example_3srcs} and \ref{f:Example_5srcs} were obtained from a single Markov chain of $5 \times 10^5 \times N$ and $15 \times 10^5 \times N$ elements, respectively; where the first $25000 \times N$ elements were discarded as a part of the burn-in period. The length of the chains is sufficient to demonstrate the ability of the DR MCMC algorithm to find multiple sources, but in a real problem, we would run longer chains in order to get accurate samplings of the posterior PDFs.

In addition of the DR parameters, see Sec.~\ref{ss:proposal_params}, many others for the MCMC algorithms need to be chosen carefully. Here we summarise the most relevant for the present discussion. First of all, we are not always updating the same number of sources. We set probabilities of updating $1, 2, \ldots, N$ sources, always selecting the closest ones in frequency, since this is the main parameter that determines the degree of correlation between two sources: two sources with exactly the same parameters but separated by, say $1000 f_m$, are perfectly resolvable. Our proposal density functions during the normal MCMC evolution are single Gaussian distributions centered at the actual element of the chain, with constant values for the amplitude of the proposals, $\Delta_i$, throughout the evolution of the chain; for the angles, the value of $\Delta_i$ was set to one-ninth of the prior range (which covered the entire physical space). For the other parameters we use $\Delta_f = 1 f_m = 3.17 \nHz$, $\Delta_{A_\cross / \A_+} = 2/3$, $\Delta_{\log \A_+} = 0.08$ and $\Delta_{\log \sigma} = 0.01$. We also consider updates of all the parameters at a time, but always working with diagonal variance-covariance matrices for the proposals. When the number of sources is greater than $1$, the DR stage is attempted with all the selected sources at a time. We account for symmetries and degeneracies of the target distribution by proposing, with a given probability, special jumps based on our prior knowledge. Most of these `quasi-equivalent' points in the parameter space are actual degeneracies of the antenna patterns that characterise the LISA response in the low frequency range \cite{Cutler:1998}, and that at higher frequencies appear as secondary maxima. When searching for more than one source, we also consider big jumps in the parameter space: single Gaussian proposals of width equal to one-third of the whole prior range for all the parameters. We found them necessary (see \eg Fig.~\ref{f:Example_5srcs}) because converged chains act like ``walls" in the frequency space, decreasing the chances of other chains to cross over them.

\section{Conclusions and future work}
\label{s:conclusions}

The motion of the LISA instrument during the observations of long lived GWs modulates the phase and amplitude of the detected signal, producing likelihood functions with a complicated structure consisting of several isolated local maxima. This is a common feature of all long lived LISA sources and, in turn, makes standard MCMC samplers inefficient. Here, we have presented a new strategy for the fully Markovian sampling of multimodal posterior PDFs based on a Delayed Rejection MCMC method and we have demonstrated its performance by applying it on the search for multiple stellar binary systems in Gaussian and stationary noise. 

The Delayed Rejection technique introduced here, can be fruitfully applied in many other LISA data analysis areas involving more complex signals, such as those from spinning massive black hole binaries and extreme-mass ratio inspirals; furthermore, it needs to be extended to the case of trans-dimensional problems, relevant to the analysis of LISA data sets to search for the whole galactic populations of stellar-mass binary systems, where the number of GW sources in unknown. In fact, an additional benefit of having a fully Markovian algorithm capable of efficiently exploring the target distribution is the possibility of estimating the Bayes factor of different models directly from the output chains. We are currently exploring the possibility of combining the Delayed Rejection technique with a Reversible Jump MCMC algorithm in order to search for an unknown number of sources.

\section*{Acknowledgments}

All the data sets used in this paper were created using the LISA tools \cite{mldc_wiki, Arnaud:2006gm} and our stellar binary waveform generator is based on an implementation provided by Cornish and Littenberg \cite{Cornish:2007if}. MT acknowledges the University of Birmingham for hospitality while this work was carried out and is grateful for the support of the Spanish Ministerio de Educaci\'on y Ciencia Research Projects FPA-2007-60220, HA2007-0042, CSD207-00042 and the Govern de les Illes Balears, Conselleria d'Economia, Hisenda i Innovaci\'o.  AV and JV acknowledge the support by the UK Science and Technology Facilities Council.

%%%%%%%%%%%
%%%%%%%%%%%

\end{document}